\input epsf.tex
\documentstyle[12pt]{article}
\begin{document}
\begin{flushright}
LSUHE No. 344-2000
\end{flushright}
\def \beq{\begin{equation}}
\def \eeq{\end{equation}}
\def \ps{\psi}
\def \pb{\bar \psi}
\def \gi{\gamma_{i}}
\def \go{\gamma_{0}}
\def \g5{\gamma_{5}}
\baselineskip=24pt
\vspace{1.5cm}
\begin{center}
\bf{ $Z(2)$ vortex solution in a field theory}
\vspace{1.0cm}
\rm

Srinath Cheluvaraja \\

\it{Dept. of Physics and Astronomy, Louisiana State University,
Baton Rouge, LA, 70808} \\
\end{center}

\noindent{\bf{ABSTRACT}}\\
\rm
We present a finite energy
topological $Z(2)$ vortex solution in a $2+1$ dimensional $SO(3)$ gauge
field theory minimally coupled to a matrix valued Higgs
field. The vortex carries a $Z(2)$ magnetic charge and obeys a modulo two addition property.
The core of this vortex has a structure
similar to that of the Abrikosov vortex appearing in a type II
superconductor.
The implications of
this solution for Wilson loops are quite interesting. In two Euclidean
dimensions these vortices are instantons and
a dilute gas of such vortices disorders Wilson loops producing an
area law behaviour with an exponentially small string tension. In $2+1$
dimensions the vortices are loops and they affect the
same disordering in the phase having large loops.
\vspace{0.5cm}
\begin{flushleft}
\end{flushleft}

\newpage
\noindent
Vortices have played an important role in statistical mechanics and
field theories. They are global excitations which arise because of
the topology of the dynamical variables. Vortices play an important role
in driving the phase transition in the two dimensional 
planar model \cite{kost}. They
also appear as a result of many body interactions in a Type II superconductor
\cite{abri}. An example of a vortex solution in a quantum field theory
was given in \cite{niels}. This solution is the relativistic version of the
Abrikosov vortex in the Type II superconductor. These vortices have an
additive
integer charge associated with them. Vortices were defined for non-abelain
gauge theories in \cite{hooft}. In this definition the vortex has a
$Z(N)$ charge. The effect of such a vortex is to produce $Z(N)$ phase
factors for Wilson loops surrounding the vortex.
Vortices are just one of
the many topological solutions that occur in field theories.
Other well known examples of topological solutions are,
the 't Hooft-Polyakov monopole \cite{mono}, the instanton \cite{inst}, and
the skyrmion \cite{skyr}. These solutions have
played an important role in uncovering the non-perturbative effects in
quantum field theories. Since they are global
effects they have consequences which cannot be seen in the usual perturbative
expansion. The 't Hooft-Polyakov monopole solution was used in \cite{poly}
to study the Higgs phase of the Georgi-Glashow model and it was shown it
radically alters the naive spectrum of the theory. Instanton solutions have been
used with varying success \cite{inphys} to understand the long distance
properties of $QCD$. Skyrmion solutions \cite{skyr} also lead to important
non-perturbative effects.
The vortex
solutions considered so far were abelian vortices, they have an integral
vortex charge. We present here an example of a $Z(2)$ vortex solution
and then examine the consequences of this solution for the field theories
in which it occurs.

Since we will be interested in vortex solutions we mention some general
properties of vortices, first in three and then in four space-time
dimensions. In three dimensions a $Z(2)$ 
vortex is said to pierce a two dimensional region 
(simply connected) $R$ if every Wilson loop
surrounding this region picks up a phase ($Z(N)$ for the group $SU(N)$).
The effect of this vortex can also be understood as the action of a pure gauge transormation
which is not single valued on every closed loop surrounding $R$. Because of
topological considerations the region $R$ cannot be arbitrarily shrunk by
regular gauge transformations and this is the region associated with the
core of a vortex. If the vortex solution is to have a finite energy the
region outside the core must have a zero energy density.
 The existence of vortex
solutions is determined by the first homotopy group, $\pi_{1}(H)$, of the field
configurations in the region oustide the vortex ($H$ being
the field configuration space outside the vortex).
The vortex can extend in the dimensions orthogonal to $R$ and
can either stretch indefinitely, form closed loops, or end in objects( monopoles) which absorb the vortex flux.
In four dimensions the above picture gets repeated on every
slice in the extra dimension and the vortex line becomes a vortex sheet whose
area is the product of the length of the vortex and the duration in time for
which the vortex propagates. These vortex sheets can form closed two
dimensional surfaces or they can end in monopole loops.
A famous example of a vortex solution in a gauge
theory is the Nielsen-Olesen vortex which occurs in three dimensional 
scalar QED \cite{niels}.
This vortex carries an integral magnetic flux
(because $\pi_{1}(U(1))=Z$), and it is a relativistic generalization
of the Abrikosov vortex appearing in a Type II superconductor.
Apart from the Abrikosov-Nielsen-Olesen vortex there are not
many other vortex solutions known.
We present here an example of a vortex solution where the vorticity can
have only one non-trivial value. This solution occurs in an $SO(3)$ invariant gauge
theory coupled to a Higgs field.

The field theory under consideration is an $SO(3)$ gauge invariant
theory minimally coupled to a matrix valued Higgs field $M$ in the $(3,3)$
representation of $SO(3)$.
The Lagrangian (in $2+1$ space-time dimensions) is given by
\beq
L= \int d^{2}x\ d\tau\frac{1}{2} tr((D_{\mu}M)^{t}(D_{\mu}M)) -\frac{1}{4}F_{\mu \nu}^{\alpha}
F_{\mu \nu}^{\alpha} -tr V((M^{t}M))
\quad .
\eeq
$D_{\mu}M$ is the covariant derivative and is defined in the usual way as
$D_{\mu}M=\partial_{\mu}M+g [A_{\mu},M]$. The gauge field $A_{\mu}$ can be
written in terms of its Lie Algebra valued components as
$A_{\mu}=A_{\mu}^{\alpha}T^{\alpha}$. $T^{\alpha}$ are $3X3$ matrices in
the Lie algebra of $SO(3)$ and satisfy $[T^{\alpha},T^{\beta}]=\epsilon^{\alpha
\beta \gamma}T^{\gamma}$ (note that since we are dealing with $SO(3)$ there
is no $i$ factor). Under local $SO(3)$ transformations ($V(x)$) 
the fields transform as
\begin{eqnarray*}
M\rightarrow VMV^{-1} \\
A_{\mu} \rightarrow V A_{\mu} V^{-1} -\frac{1}{g}(\partial_{\mu}V)V^{-1}
\quad .
\end{eqnarray*}
The equations of motion are given by
\begin{eqnarray*}
D_{\mu}F_{\mu \nu \alpha} =-g\ tr((D_{\nu}M)^{t}[T^{\alpha}_{\nu},M]) \\
\partial_{\mu}(D_{\mu}M)=-[A_{\mu},D_{\mu}M] -\frac{\partial V}{\partial M}
\quad .
\end{eqnarray*}
(There is no summation over the $\nu$ index in the rhs of the first equation).
We will be interested in classical vortex
solutions of these equations of motion.
We mimic the approach taken in \cite{niels} to look for these solutions.
Outside the core of the vortex we require the fields to have zero energy density
so that the solution has a finite energy. A set of such zero
energy density configurations outside the core must satisfy
\begin{eqnarray*}
D_{\mu}M=0 \\
F_{\mu \nu}^{\alpha}=0 \\
V(M)=0 \\
\frac{\partial V}{\partial M}=0
\quad .
\end{eqnarray*}
The last two conditions ensure that the potential energy density (associated
with the field M) is zero outside the vortex core and that the field M
is always at the minimum of the potential energy (also taken to be zero).
A field obeying these conditions is said to be in a Higgs vacuum.
A vortex solution is one which winds around in a non-trivial way in
some internal space for a closed loop traversal in physical space. It is clear that
the existence of such configurations is determined by the structure of
the first homotopy group of the Higgs vacuum. The Abrikosov-Nielsen-Olesen
vortex is an example of a realization of a non-trivial element
the first Homotopy group of the
group $U(1)$. The relevant group in our example is $SO(3)$.
The Higgs vacuum can be chosen in such a way that it consists of elements
of SO(3). One choice of Higgs potential that accomplishes this is
\beq
V(M)= \lambda (M^{t}M-I)^2
\quad .
\eeq
The minimum of this potential is the set of orthogonal matrices
\beq
M^{t}M=I
\quad .
\eeq
Any choice of the vacuum $M$ satisfying the above condition breaks the
$SO(3)$ gauge symmetry. However, a specific constant choice of $M$, say $M_{s}$,
is still invariant under a subgroup of $SO(3)$ gauge transformations
such that
\beq
[R,M_{s}]=0
\quad .
\eeq
Thus there is still an unbroken gauge symmetry corresponding to the
$SO(3)$ matrices which commute with $M_{s}$. The set of such
matrices form a subgroup $SO(2)$ of $SO(3)$.
If the matrix $M_{s}$ is chosen to be
\beq
M_{s}=\exp (T^{3}\alpha)
\eeq
for some $\alpha$, then gauge transformations about the $T^{3}$ axis
will leave the vacuum invariant. The field $A_{\mu}^{3}$ will be the
massless $SO(2)$ gauge field.
Excitations about $M_{s}$
can be described by a gauge theory (with gauge group $SO(2)$) in which there
are two oppositely charged massive gauge bosons, one massless boson, and a scalar field.
However, we are
not interested in the excitations about
the conventional Higgs phase. We are looking for global
solutions which, even though they belong to the Higgs vacuum, cannot be
obtained continuously (to any finite order in perturbation theory)
from the conventional Higgs phase.
We look for static solutions having zero energy density 
outside the core of the vortex.
We also impose the
following conditions
\begin{eqnarray*}
A_{0}^{\alpha}=0 \\
A_{\mu}^{1}=A_{\mu}^{2}=0
\quad .
\end{eqnarray*}
The first of the above conditions, along with the static condition,
implies a zero electric field. In addition to these conditions,
only the $3$ component (of the Lie algebra) of the gauge field is taken to be 
non-zero and this is
chosen to have the same form as in the A-N-O vortex
\begin{eqnarray*}
A_{1}^{3}(x)=\frac{-y}{g r^2} \\
A_{2}^{3}(x)=\frac{x}{g r^2}
\quad .
\end{eqnarray*}
In polar co-ordinates the solution has the form
\begin{eqnarray*}
A_{\theta}^{3}=\frac{1}{g r} \\
A_{r}^{3}=0
\quad .
\end{eqnarray*}
The above solution satisfies $F_{12}^{3}=0$. We note for later reference
that any $A_{\theta}^{3}$ proportional to $\frac{1}{r}$ is also a
solution.
The field  equation for $M$ is satisfied if we take
$M$ to be a independent of $r$ and $M^{t}M=I$. This leads to the
following equation for $M$
\beq
\frac{1}{r} \frac{\partial M}{\partial \theta} +g [A_{\theta},M]=0
\quad .
\eeq
Since only the $3$ component of $A_{\mu}$ is non-zero the above
equation  reduces to
\beq
\frac{1}{r} \frac{\partial M}{\partial \theta} +g A_{\theta}^{3} [T^{3},M]=0
\quad .
\label{covdar}
\eeq
The above solution can be solved easily by noting that for small
$\delta \theta$,
\beq
M(\theta +\delta \theta)=\exp (-T^{3}\delta \theta) M(\theta) \exp(T^{3}\delta \theta)
\eeq
solves Eq.~\ref{covdar} for $A_{\theta}^{3}=\frac{1}{g r}$.
This leads to the following solution
\beq
M_{0}(\theta)=\exp (-T^{3}\theta) M(0) \exp(T^{3}\theta)
\quad .
\eeq
The gauge field for this solution will be labelled as
$A_{\theta \ 0}^{3}$.
However, there is another solution of the form
\beq
M_{1}(\theta)=\exp (-T^{3}\frac{\theta}{2}) M(0) \exp(T^{3}\frac{\theta}{2})
\eeq
that also solves Eq.~\ref{covdar} provided $A_{\theta}^{3}=\frac{1}{2 g r}$.
This gauge field will be labelled as $A_{\theta \  1}^{3}=\frac{1}{2 g r}$.
We will show that the two solutions,$M_{0}$ and $M_{1}$,
belong to different elements of the
first Homotopy group of $SO(3)$ ($\Pi_{1}(SO(3))=Z(2)$).
To see this we need a particular parametrization of the elements of the
group $SO(3)$. An element of $SO(3)$ ($R$) 
can be parametrized as
\beq
R(n,\tilde \theta)=\exp (T^{\alpha}n^{\alpha}\tilde \theta)
\quad ,
\label{param}
\eeq
where $n$ is a unit vector which takes values on the surface of a
two dimensional sphere. $\tilde \theta$ varies in the range $0 \le \tilde \theta \le \pi$.
Every $SO(3)$ element is parametrized by an axis of rotation
$n$ and an angle of rotation $\tilde \theta$.
The $SO(3)$ manifold is thus the
solid ball of radius $\pi$.
Since a rotation around an axis $n$ by an angle $\pi$
is the same as the rotation around the axis $-n$ by the same angle $\pi$, 
antipodal points on
the surface of the ball are identified. It is this
identification that gives a non-trivial topology to the $SO(3)$ group.
The matrix $R$ can be written out as
\beq
R(n,\tilde \theta)=I+(T^{\alpha}n^{\alpha})\sin(\tilde \theta) +(T^{\alpha}
n^{\alpha})^{2}(1-\cos(\tilde \theta)) \quad .
\eeq
(We have used the property $(T^{\alpha})^{3}=-T^{\alpha}$,
expanded the exponential,
and reordered the various terms.)
The matrices $T^{\alpha}$ are the antisymmetric $3X3$ matrices
belonging to the Lie algebra of $SO(3)$.
Of the two solutions presented earlier,
one is topologically trivial,
the other is topologically non-trivial and this is the one that will
be referred to as the vortex.
The solutions presented are
just one of a class of many possible solutions.
In the above solutions if we choose $M(0)=\exp T^{2}\pi$, it is easy to
see that $M_{1}(\theta)$ moves along the equator of the ball and returns to
itself whereas $M_{2}(\theta)$ goes to the point diametrically opposite to
$M_{0}$.
This follows from the property
\beq
\exp(T^{3}\theta)\exp(T^{2}\pi)\exp(-T^{3}\theta)=\exp(T^{\alpha}n^{\alpha}\pi)
\eeq
where $n^{1}=-\sin\theta \ \ n^{2}=\cos\theta \ \ n^{3}=0$.
The way of visualizing this is that the point on the surface of the ball
$\tilde \theta=\pi,n^{1}=n^{3}=0,n^{2}=1$ gets rotated in the equatorial plane
of the ball ($n^{1}-n^{2}\ plane$) by an angle $\theta$. In the solution
$M_{0}$, the vector $n$ rotates by $2\pi$
whereas in the solution $M_{1}$ the
vector $n$ rotates by $\pi$, and the two paths belong to different elements
of $\Pi_{1}(SO(3))$. Since opposite points are identified only
on the surface of the ball,
it is crucial to choose $M(0)$ to lie on the surface of the
ball in the solution $M_{1}(\theta)$. In $M_{0}(\theta)$ the point $M(0)$
can be chosen anywhere, although it has also been chosen to be on the
surface of the ball,  because this is the solution which is topologically
trivial.
The two solutions can be transformed into each other by the following singular gauge
transformation
\beq
V_{s}(\theta)=\exp(-T^{3}\frac{\theta}{2})
\quad .
\eeq
This follows from the relation
\beq
M_{0}(\theta)=V_{s}(\theta)M_{1}(\theta)V_{s}^{-1}
\quad .
\eeq
This gauge transformation takes $A_{\theta \ 1}^{3}=\frac{1}{2 g r}$ to
$A_{\theta \ 0}^{3}=\frac{1}{g r}$.
The above gauge transformation has a discontinuity on the $\theta =0$ axis because
\beq
V_{s}(2\pi)=\exp(-T^{3}\pi)\ne V_{s}(0)
\quad .
\eeq
($Tr V_{s}(2\pi)=-1,\ TrV_{s}(0)=3$).
An
interesting feature is that two topologically non-trivial solutions
annihilate each other because
\beq
A_{\theta \  0}=A_{\theta \ 1} + A_{\theta \ 1}
\quad ,
\eeq
or, in terms of the $M$ fields,
\beq
M_{1}(\theta)^{2}=I
\quad .
\eeq
Therefore, these vortices have a $Z(2)$ magnetic charge.
It is worth pointing out that the $Z(2)$ magnetic charge that arises
here has nothing to do with the original gauge group $SO(3)$.
Unlike $SU(2)$ which has $Z(2)$ as its center, the group $SO(3)$ has
no non-trivial center subgroup.
We note that the solution $M_{0}(\theta)$ can be completely gauge
transformed away and corresponds to the zero field case (and hence zero energy) case.
The above solutions are valid only outside the vortex core; in order
to have a finite energy non-singular solution inside the vortex core
we must look for solutions
which are not singular at the core and which continuously go over to the
solutions outside the core.
We look for solutions inside the core by writing 
\begin{eqnarray*}
A_{\theta}^{3}(r)=A_{\theta 1}^{3}(r) +A_{\theta c}^{3}(r), \\
M(r, \theta)=M_{1}(\theta)M_{c}(r)
\quad .
\end{eqnarray*}
It is assumed above that the angular behaviour of the $M$
field inside the vortex core is the same as that outside the core. The
only additional piece in $M$
is a radial dependence $M_{c}(r)$. The matrix $M_{c}(r)$
is a multiple of the identity and can be moved through the
matrix $M_{1}(\theta)$. There is also an additional piece for the
gauge field, $A_{\theta c}$, which also  contributes to the energy density of the vortex.
Substituting the above equations in the field
equations we get the following differential equations for $A_{\theta c}$ and
$M_{c}(r)$.
\begin{eqnarray*}
-\frac{d^{2}M_{c}(r)}{dr^{2}} -\frac{d M_{c}(r)}{r\ dr} =
g^{2} (A_{\theta c})^{2}(r)
([T^{3},[T^{3},M(0)]] M(0)^{-1}) -\frac{d\bar V(M)}{dM}|_{M=M(0)}{M(0)^{-1}}
\\
\frac{d^{2}A_{\theta c}^{3}}{dr^{2}} +\frac{d}{dr}(\frac{A_{\theta c}^{3}}{r})=
-g M_{c}^{2}(r) A_{\theta c}^{3}(r) tr ([T^{3},M(0)]^{t}[T^{3},M(0)])
\quad .
\end{eqnarray*}
\beq
\frac{d\bar V(M)}{dM}|_{M=M(0)}M(0)^{-1}=\lambda (M_{c}(r)^2-I)
\eeq
The above differential equations have the same form as those of the
A-N-O vortex (the matrices can be traced over and yield only
some constant factors).
Although this is not very surprising, because we are looking at
abelian field configurations (only $A_{\mu}^{3} \ne 0$), it is interesting
to note that they define the core of a vortex which is quite different
from the A-N-O vortex. This difference  arises because the abelian group $SO(2)$
is now embedded in $SO(3)$.
The differential equations for the vortex core are quite complicated and
cannot be solved analytically. For regions far from the center of the vortex
(but still inside the core of the vortex)
we can use the following approximate solutions for $M_{c}(r)$ and
$A_{\theta c}(r)$.
\begin{eqnarray*}
A_{\theta c}(r)=(const) \frac{\exp(- \sqrt {8} g r)}{\sqrt r } \\
M_{c}(r)=1-\exp(- \sqrt {\lambda} r)
\quad .
\end{eqnarray*}
The implications of such vortex solutions for Wilson loops are
quite striking.
Unlike the A-N-O vortex, which always contributes
a factor of $+1$ to the Wilson loop, or the monopole, which produces
a phase depending on the solid angle subtended at it by the Wilson loop,
the $Z(2)$ vortex produces a constant factor of $-1$ whenever it
lies inside the minimal surface spanned by the Wilson loop.
This effect can be demonstrated by looking
at
\beq
\exp g \int_{C}  A_{\mu}(x)dx^{\mu}
\quad .
\eeq
If the loop $C$ encircles a vortex ( and has a size larger than the
core of the vortex) the Wilson loop is either
$\exp 2\pi T^{3}$ (for $A_{\theta 0}$)
or $\exp \pi T^{3}$ (for $A_{\theta 1}$). The trace of
the first quantity is $3$ whereas the trace of the second quantity is $-1$
(see Eq.~\ref{param}).
These vortices have a $Z(2)$ magnetic charge and, as far as the Wilson
loop is concerned,
only their total
parity (odd or even number) is physical, their overall number being
irrelevant.
The energy of the vortex is finite and
can be written as
\beq
E=E_{1}(\lambda)/g^2
\quad ,
\eeq
where $E_{1}(\lambda)$ is a quantity which depends on the parameters in the
Higgs potential. The size of the vortex is determined by the parameter
$\lambda$  ($\frac{1}{\sqrt \lambda}$).
The energy cannot be determined exactly in the absence of an
exact solution inside the vortex core.
We can also construct approximate multi-vortex solutions by superposing many single vortex
solutions and making their mutual separations much larger than their
individual sizes.

If we now consider the system in $2$ Euclidean dimensions the vortex solutions
have finite action and they can be used as a starting point for a semi-classical
approximation. This is a well known scheme for approximating the partition
function and was brought to its culmination in \cite{poly} where the
three-dimensional Georgi-Glashow model was analyzed in the
background of 't Hooft-Polyakov monopoles and
shown to be in the confining phase. The semi-classical technique proceeds
by expanding the partition function about the classical vortex solution
by writing
\begin{eqnarray*}
A_{\mu}(x)=\bar A_{\mu}(x) + a_{\mu}(x) \\
M(x)=\bar M(x)\ m(x)
\quad .
\end{eqnarray*}
Here we denote $\bar A_{\mu}$ and $\bar M(x)$ as the classical solutions
and $a_{\mu}(x)$ and $m(x)$ as the quantum fluctuations
about the classical solution. The partition function can be expanded as
\beq
Z= \int DA_{\mu} DM(r) \exp -(S_{c} + \frac{\delta^{2}S_{c}}{2 \delta a_{\mu}(x)
\delta a_{\nu}(y)} a_{\mu}(x)a_{\nu}(y)
+\frac{\delta^{2}S_{c}}{2 \delta \bar M(x)\delta \bar M(y)}tr (m^{t}(x)m(y))
+ \ \ ...)
\quad .
\eeq
($S_{c}$ is the action of the vortex solution) The semi-classical method is
a good approximation provided we are in the weak coupling regime.
If we simply consider the classical solution and ignore the quantum
fluctuations the expectation value of the Wilson loop of area $A$ becomes
\beq
\frac{1}{Z} \sum_{n=0}^{\infty} \sum_{n_{1}=0}^{n}
n\ C_{n_{1}} \exp -(n E_{1}/(g^2)) (-1)^{n_{1}} (\frac{A
}{\Omega})^{n_{1}} (1-\frac{A}{\Omega})^{n-n_{1}}
\quad .
\eeq
In the above expression, $(-1)^{n_{1}}$ is the effect on the Wilson loop
if it encloses $n_{1}$ vortices,$\frac{A}{\Omega}$ is the probability of
having a vortex enclosed by the Wilson loop, $(1-\frac{A}{\Omega})$ is the
probability of the vortex not enclosing the Wilson loop,$n\ E_{1}$ is the
energy associated with $n$ vortices with their mutual interactions
ignored, $n$ is the total number of vortices in the system, $\Omega$ is the
spatial area of the system.
The major approximation made is the neglect of vortex-vortex interactions
(also called the dilute gas approximation)
and the construction of multi-vortex solutions by superposing single vortex
solutions with large mutual separations ($>> \frac{1}{\sqrt \lambda}$).
The sum can be easily evaluated using the properties of
the binomial expansion and it becomes
\beq
Lim_{n \rightarrow \infty}(1-\frac{2 A \rho}{n} \exp (-\frac{E_{1}}{g^2}))^n
\quad ,
\eeq
giving an exponentially small string tension
\beq
\sigma =2 \rho \exp(\frac{-E_{1}}{g^2})
\quad .
\eeq
($\rho$ is the density of vortices in the system). The
density of vortices is proportional to $\exp -\frac{E_{1}(\lambda)}{g^2}$
and is very small in the weak coupling region. We are thus consistent in
using the dilute vortex approximation.
The calculation of the quantum fluctuations will require the calculation
of determinants of second order differential operators in the background
of the vortex solution. Fluctuations of the vortex which are just
translations or gauge rotations will lead to zero modes in the
determinant.
The zero-modes that appear in this integration
will correspond to the translational(2) and gauge invariance(3) of the vortex
solution. The zero modes can be handled by the collective co-ordinate
method and it is well known
 \cite{raj} that integration over the zero modes 
gives an additional
correction proportional to
\beq
g^{-n_{v}}
\eeq
where $n_{v}$ is the number of zero modes. The numerical factor that
accompanies this requires a very detailed calculation of the
determinant operator and will not be calculated here. We can
write the result of the first-order quantum fluctuations as
\beq
g^{-5}\epsilon(\lambda)
\quad .
\eeq
$\epsilon(\lambda)$ is the result of the calculation of the determinant
operator (without the zero modes) in the background of the vortex solution. 
The effect of this calculation is to renormaize the chemical potential of the
vortices to
\beq
\frac{1}{g^{5}} \epsilon(\lambda) \exp \frac{-E_{1}(\lambda)}{g^2}
\quad .
\eeq

In the $2+1$ dimensional field theory 
the vortices were presented as static solutions
but Lorentz boosts of these solutions are also solutions-
the vortices form
loops which can be thought of as the world-lines of particles carrying a $Z(2)$
magnetic charge. 
The energy of these
loops will be proportional to their length (there will also be factors coming
from the thickness, $\frac{1}{\sqrt \lambda}$).
 Such a dilute vortex gas
can either form a entropy dominated vortex condensate with long vortex loops,
or a energy dominated phase in which the vortex loops are very
small.
Now the vortex produces a negative sign
for a Wilson loop provided it has a non-trivial linkage with it. Large
vortex loops will be necessary if they have to produce the 
disordering (of of the kind just exhibited in 2 dimensions) of large Wilson loops,
and this is only possible in the phase in which the vortices condense.
In the phase with small vortices large Wilson loops will not be disordered likewise
and only a perimeter law will result.
At weak coupling, where the semi-classical method is applicable, the vortex
energy is very large and only small vortices are formed. However, at strong
coupling the vortex entropy dominates and large vortices will be present.

\noindent
Acknowledgement: This work was supported in part by United States Department
of Energy grant DE-FG 05-91 ER 40617.

\end{document}